\documentclass[12pt]{article}

\usepackage{amsfonts,amsmath,amssymb,accents,mathrsfs}
\usepackage[retainorgcmds]{IEEEtrantools}
\usepackage{multirow}
\usepackage{multicol}
\usepackage{tikz}
\usepackage{graphicx,color}
\usepackage{booktabs}
\usepackage{placeins}
\usepackage[colorlinks,linkcolor=Blue,citecolor=Blue,bookmarks,bookmarksnumbered]{hyperref}
\usepackage{cite}
\usepackage[scaled=0.85]{helvet}
\usepackage[utf8]{inputenc}
\usepackage{todonotes}
\usepackage{color,soul}

\definecolor{Red}    {rgb}{0.90,0.00,0.12} %  1
\definecolor{Blue}   {rgb}{0.00,0.00,1.00} %  2
\definecolor{Green}  {rgb}{0.10,0.70,0.10} %  3
\definecolor{Turque} {rgb}{0.00,0.65,0.85} %  4
\definecolor{Orange} {rgb}{1.00,0.50,0.15} %  5
\definecolor{Magenta}{rgb}{1.00,0.00,1.00} %  6
\definecolor{Gold}   {rgb}{1.00,0.75,0.25} %  7
\definecolor{Seaweed}{rgb}{0.01,0.24,0.09} %  8
\definecolor{Purple} {rgb}{0.50,0.25,0.55} %  9
\definecolor{Brown}  {rgb}{0.43,0.26,0.32} % 10
\definecolor{grey1}  {rgb}{0.20,0.20,0.20} % 11
\definecolor{grey2}  {rgb}{0.40,0.40,0.40} % 12
\definecolor{grey3}  {rgb}{0.60,0.60,0.60} % 13
\definecolor{grey4}  {rgb}{0.80,0.80,0.80} % 14
\definecolor{grey5}  {rgb}{0.90,0.90,0.90} % 15

\def\a{{\alpha}}
\def\b{{\beta}}
\def\g{{\gamma}}
\def\d{{\delta}}

\def\s{{\sigma}}

\def\th{{\theta}}
\def\S{{\Sigma}}

\def\ad{{\dot{\alpha}}}
\def\bd{{\dot{\beta}}}
\def\gd{{\dot{\gamma}}}

\def\thd{{\bar{\theta}}}
\def\Sd{{\bar{\Sigma}}}

\def\A{{\mathcal{A}}}

\def\N{{\mathcal{N}}}

\def\J{{\mathcal{J}}}
\def\T{{\mathcal{T}}}
\def\Y{{\mathcal{Y}}}

\def\D{{\rm D}}
\def\Dd{{\bar{\rm D}}}
\def\pa{\partial}

\def\be{\begin{equation}}
\def\ee{\end{equation}}
\def\bea{\begin{IEEEeqnarray*}}
\def\eea{\end{IEEEeqnarray*}}
\def\n{\IEEEyesnumber}
\def\sn{\IEEEyessubnumber}

\makeatletter
\def\section{\@startsection{section}{1}{\z@}
              {3ex plus-1ex minus-.2ex}{1pt plus1pt}{\large\sf\bfseries\boldmath}}
\def\subsection{\@startsection{subsection}{2}{\z@}
              {1.5ex plus-1ex minus-.2ex}{0.01pt plus1pt}{\sf\slshape}}
\def\subsubsection{\@startsection{subsubsection}{3}{\z@}
              {1.5ex plus-1ex minus-.2ex}{0.01pt plus0.2pt}{\sf\boldmath}}
\def\paragraph{\@startsection{paragraph}{4}{\z@}
              {.75ex \@plus.5ex \@minus.2ex}{-2mm}{\sf\bfseries\boldmath}}
\makeatother

   \parskip\medskipamount     \lineskip=0pt
\thicklines
\begin{document}
\thispagestyle{empty}
\noindent{\small
%\hfill{HET- {~} \\ % un-comment-out and specify when done}
$~~~~~~~~~~~~~~~~~~~~~~~~~~~~~~~~~~~~~~~~~~~~~~~~~~~~~~~~~~~~$
$~~~~~~~~~~~~~~~~~~~~\,~~~~~~~~~~~~~~~~~~~~~~~~~\,~~~~~~~~~~~~~~~~$
 {~}
}
\vspace*{8mm}
\begin{center}
{\large \sf\bfseries\boldmath  %\bf
Higher Spin Superfield interactions with Complex linear Supermultiplet:
\\ [2pt]
Conserved Supercurrents and Cubic Vertices } \\   [12mm] {\large {
Konstantinos Koutrolikos\footnote{kkoutrolikos@physics.muni.cz},
Pavel Ko\v{c}\'{i}\footnote{pavelkoci@mail.muni.cz} and
Rikard von Unge\footnote{unge@physics.muni.cz}}}
\\*[10mm]
\emph{
\centering
Institute for Theoretical Physics and Astrophysics, Masaryk University,
\\[1pt]
611 37 Brno, Czech Republic
}
 $$~~$$
  $$~~$$
 \\*[-8mm]
{ ABSTRACT}\\[4mm]
\parbox{142mm}{\parindent=2pc\indent\baselineskip=14pt plus1pt
We continue the program of constructing cubic interactions between matter 
and higher spin supermultiplets.
In this work we consider a complex linear superfield and we find that it can 
have cubic interactions only with
supermultiplets with propagating spins $j=s+1$, $j=s+1/2$ for any non-negative 
integer $s$ (half-integer superspin supermultiplets).
We construct the higher spin supercurrent and supertrace, these compose the 
canonical
supercurrent multiplet which generates the cubic interactions. We also prove 
that for every $s$ there exist an alternative minimal
supercurrent multiplet, with vanishing supertrace. Furthermore, we perform a 
duality transformation
in order to make contact with the corresponding chiral theory. An interesting 
result is that the dual chiral theory has the same
coupling constant with the complex linear theory only for odd values of $s$, 
whereas for even values of $s$ the coupling constants
for the two theories have opposite signs. Additionally we explore the component 
structure of the supercurrent multiplet and derive the higher spin currents. We 
find two bosonic currents for spins $j=s$ and $j=s+1$ and one fermionic current 
for spin $j=s+1/2$.
}
 \end{center}
$$~~$$
\vfill
%\noindent PACS: 11.30.Pb, 12.60.Jv\\
\vfill
\clearpage
%
%%%%%%%%%%%%%%%%%%%%%%%%%%%%%%%%%%%%%%%%%%%%%%
%%%%%%%%% INTRODUCTION %%%%%%%%%%%%%%%%%%%%%%%
\section{Introduction}

In a recent paper \cite{hsch1} the cubic interactions between the chiral 
supermultiplet and higher spin
supermultiplets were constructed via Noether's procedure. The conclusion was 
that 
both massive and massless chiral superfields
couple only to half-integer superspin supermultiplets $(s+1 , s+1/2)$ with an 
interesting twist that the integer $s$ must be odd for the
case of massive chirals. Furthermore, excplicit expressions were given for the 
supercurrent multiplet which includes the higher spin supercurrent
$\J_{\a(s)\ad(s)}$\footnote{The notation $\a(s)$ represents a string of $s$ undotted, symmetric indices $\a$, $\a_1\a_2\dots\a_{s}$. } and the higher spin supertrace $\T_{\a(s-1)\ad(s-1)}$.

This paper is the continuation of that program for a different matter 
supermultiplet. In this work, the role of matter will be played by the
complex linear supermultiplet and the goal is to (\emph{i}) find the spin 
selection 
rules for consistent cubic interactions between the 
complex linear and the $4D,~\N=1$ super-Poincar\'{e} higher spin supermultiplets 
and 
(\emph{ii}) give explicit expressions for the higher spin supercurrent 
supermultiplets.

The existence of \emph{variant} descriptions of supersymmetric theories has been 
well documented \cite{variant} and one of the most 
well known examples are the minimal and non-minimal formulations of $4D$, 
$\N=1$ supergravity. The complex linear supermultiplet is
another well known, non-minimal, description of the scalar multiplet. In a 
previous paper \cite{cls} we derived the supercurrent multiplets
for various free and interacting (including higher derivatives) theories of a 
complex linear superfield ($\S$) by investigating the linearized coupling to 
supergravity generated by linearized superdiffeomorphisms. 

In this paper, we generalize the linearized superdiffeomorphism transformation 
of $\S$ in order to 
include higher rank parameters in a manner consistent with the linearity 
constraint of $\S$ ($\Dd^2\S=0$). Using this \emph{higher spin} transformation 
of $\Sigma$, we perform a perturbative Noether's procedure in order to construct 
an invariant theory, up to first order in coupling constant. This will 
reveal 
the cubic interactions of $\S$ with the $4D,~\N=1$ super-Poincar\'{e} higher 
spin supermultiplets. From that action we can read off the corresponding 
higher spin supercurrent multiplet which generates the cubic vertices.

The results we find are that a complex linear superfield can have cubic 
interactions only with half-integer superspin supermultiplets $(s+1,s+1/2)$ 
exactly like a chiral superfield. Moreover, we provide explicit expressions for the
both the \emph{canonical} and \emph{minimal} supercurrent multiplets.
Additionally, we investigate the duality transformation
between the complex linear and the chiral in the presence of these higher
spin cubic interactions. Interestingly enough we find that 
the \emph{charge} (the coupling constant that controls these cubic 
interactions) of the dual theory differs from the charge
of complex linear by a sign that depends on the value of $s$.  Specifically, for 
even values of $s$ they are opposite and for odd values of $s$ are the same. We 
also calculate the set of component higher spin currents. There are two bosonic 
conserved currents, one for spin $j=s$ and one for spin $j=s+1$ and there is a 
fermionic current for spin $j=s+1/2$. These currents agree with the expressions
derived from the chiral theory presented in \cite{hsch1}.

The paper is organized in the following manner. In \textsection\ref{dSigma} we 
consider a family of first order transformations for the complex linear 
superfield and their compatibility with the linearity condition. In 
\textsection\ref{CM} we use them for Noether's procedure 
in order to construct an invariant action. This will force us to introduce the 
cubic interactions between $\S$ and higher spin. These interactions are 
generated by the \emph{canonical} higher spin supercurrent multiplet which 
includes the higher spin supercurrent and supertrace.  In \textsection\ref{MMM} 
we prove that always exist an alternative \emph{minimal} supercurrent multiplet 
in which the higher spin supertrace vanishes. This is possible due to (\emph{i}) the 
freedom of choosing appropriate improvement terms, (\emph{ii}) the freedom to absorb 
trivial terms by redefining $\S$ and (\emph{iii}) the freedom in defining the 
supercurrent and supertrace up to a particular equivalence relation. In 
\textsection\ref{SecQE} we discuss the superspace conservation equations for the 
supercurrent multiplets and in \textsection\ref{Duality} we compare the results 
found for the complex linear with the results for chiral, by performing the 
duality between them. In \textsection\ref{SecComp} we extract the component 
higher spin currents and in \textsection\ref{SecConclude} we summarize our 
results.

For additional developments on the topic of supersymmetric higher spin supercurrents the reader can
refer to \cite{HK1,HK2}.

%%%%%%%%%%%%%%%%%%%%%%%%%%%%%%%%%%%%%%
%%%%%%%%%%%%%% delta \Phi %%%%%%%%%%%%%%%%%%%
\section{First order transformation for complex linear superfield}\label{dSigma}
A cubic interaction between two types of fields can be
written in the form $jh,$ where $j$ is a current constructed from
matter fields $\phi$ and $h$ is a set of gauge fields. Because the gauge
field $h$ is defined up to a gauge redundancy, the current $j$
must be conserved.
Noether's method is a systematic and perturbative method for constructing
invariant theories that respect these gauge redundancies and therefore generate
the appropriate interactions between matter and gauge fields.
In this approach one expands the action $S[
\phi,h]$ and the transformation of fields in a power series of a coupling
constant $g$
\bea{l}
S[\phi,h]=S_0[\phi]+gS_1[\phi,h]+g^2S_2[\phi,h]+\dots\n~,~\\
\delta \phi=\delta_0[\xi]+g\delta_1[\phi,\xi]+g^2\delta_2[\phi,\xi]+\dots\n~,~\\
\delta h=\delta_0[\zeta]+g\delta_1[h,\zeta]+g^2\delta_2[h,\zeta]+\dots\n
\eea
where $S_i[\phi,h]$ includes the interaction terms of order $i+2$ in the
number of fields and $\delta_i$ is the part of transformation with terms
of order $i$ in the number of fields. The requirement for invariance up to order 
$g^1$
(cubic interactions) is
\bea{l} \n\label{cub-inv-constr}
g\frac{\delta S_0}{\delta\phi}\delta_1\phi+g\frac{\delta S_1}{\delta
\phi}\delta_0\phi
+g\frac{\delta S_0}{\delta h}\delta_1 h+g\frac{\delta S_1}{\delta h}\delta_0 h=0~.
\eea

In this work, the role of matter will be played by the complex linear 
supermultiplet,
described by a complex linear superfield $\S$ ($\Dd^2{\S}=0$).
For the role of gauge fields we consider the massless, higher spin,
irreducible representations of the $4D,~\mathcal{N}=1$,
super-Poincar\'{e} algebra. These were first introduced in \cite{hsh5,Vas} and later given
in a superspace formulation \cite{hss1,hss2,hss3} and further developed in 
\cite{hss4,hss5,hssBRST,hss6}.
Higher spin supermultiplets are parametrized by a quantum number called \emph{superspin} $Y$,
which is a supersymmetric extension of spin and it takes integer $s$ and half-integer $s+1/2$ values.
Massless higher spin supermultiplets contain two irreducible representations of the Poincar\'{e} algebra with
spins $j$ and $j-1/2$, therefore we denote the entire supermultiplet by 
$(j,j-1/2)$. We briefly remind the reader of the various superspace descriptions of
free  super-Poincar\'{e} higher spin supermultiplets:
\begin{enumerate}
\item The integer superspin $Y=s$ supermultiplets $(s+1/2 , s)$
\footnote{introduced in \cite{hss1}, however we are following the formulation given in \cite{hss5}}
are described
by a pair of superfields
$\Psi_{\ad(s)\ad(s-1)}$ and $V_{\a(s-1)\ad(s-1)}$ with the following zero
order gauge transformations
\bea{l}\n\label{hstr1}
\d_0\Psi_{\ad(s)\ad(s-1)}=-\D^2L_{\a(s)\ad(s-1)}+\tfrac{1}{(s-1)!}\Dd_{(\ad_{
s-1}}\Lambda_{\a(s)\ad(s-2)}~,\sn\\  \d_0 V_{\a(s-1)\ad(s-1)}=\D^{\a_s}L_{\a(
s)\ad(s-1)}+\Dd^{\ad_s}\bar{L}_{\a(s-1)\ad(s)}\sn~.
\eea
%%%%
\item The half-integer superspin $Y=s+1/2$ supermultiplets $(s+1 , s+1/2)$
\footnote{introduced in \cite{hss2}, however we are following the formulation given in \cite{hss5}}
have two formulations, the transverse and the longitudinal.
The transverse description uses the pair of superfields $H_{\a(s)
\ad(s)}$, $\chi_{\a(s)\ad(s-1)}$ with the following zero order gauge 
transformations
\bea{l}\n\label{hstr2}
\d_0 
H_{\a(s)\ad(s)}=\tfrac{1}{s!}\D_{(\a_s}\bar{L}_{\a(s-1))\ad(s)}-\tfrac{1}{s!}
\Dd_{(\ad_s}L_{\a(s)\ad(s-1))}\sn\label{hstr2H}~,\\
\d_0\chi_{\a(s)\ad(s-1)}=\Dd^2L_{\a(s)\ad(s-1)}+\D^{\a_{s+1}}\Lambda_{\a(
s+1)\ad(s-1)}\sn
\eea
whereas the longitudinal description uses the superfields $H_{\a(s)\ad(s)}$, $\chi_{\a(s-1)\ad(
s-2)}$ with
\bea{l}\n\label{hstr3}
\d_0 
H_{\a(s)\ad(s)}=\tfrac{1}{s!}\D_{(\a_s}\bar{L}_{\a(s-1))\ad(s)}-\tfrac{1}{s!}
\Dd_{(\ad_s}L_{\a(s)\ad(s-1))}\sn~,\\
\d_0\chi_{\a(s-1)\ad(s-2)}=\Dd^{\ad_{s-1}}\D^{\a_s}L_{\a(s)\ad(s-1)}+\tfrac{
s-1}{s}\D^{\a_s}\Dd^{\ad_{s-1}}L_{\a(s)\ad(s-1)}
+\tfrac{1}{(s-2)!}\Dd_{(\ad_{s-2}}J_{\a(s-1)\ad(s-3))}~~~~\sn
\eea
\end{enumerate}

The starting point of our analysis is the free action for a complex 
linear superfield
\bea{l}
S_{0}=-\int d^8z~\Sd~\S\n\label{freeth}
\eea
and the zeroth order transformation of $\S$, $\d_{0}\S=0$. Hence according to 
(\ref{cub-inv-constr}), the cubic interactions of the complex linear superfield with
higher spin supermultiplets, described by the $S_{1}[\S,\A]$\footnote{Where $\A$ 
is the set of superfields that participate in the description of higher spin 
supermultiplets.} must satisfy:
\bea{l}
\frac{\delta S_0}{\delta\S}\delta_1\S+\frac{\delta S_1}{\delta \A}\delta_0 \A=0\n\label{Ncon}~.
\eea
Therefore it is important to find the first order, \emph{higher spin} 
transformation of $\S$ ($\d_1\S$).
Motivated from the structure of the transformation of $\S$ under 
superdiffeomorphisms presented in \cite{cls}
we write the following ansatz\footnote{We use the conventions of 
\emph{Superspace} \cite{GGRS}, $\{\D_{\a},\Dd_{\ad}\}=i\pa_{\a\ad}$,
$\D^{\a}\D_{\a}=2\D^2$ and $\Dd^{\ad}\Dd_{\ad}=2\Dd^2$.}
\bea{ll}
\d_{1}\S=g\sum_{l=0}^{\infty}\sum_{k=0}^{\infty}&\left\{\vphantom{\frac12}
A^{\a(k+1)\ad(k)}_{l}~\Box^{l}~\D_{\a_{k+1}}\Dd_{\ad_k} 
\D_{\a_k}\dots\Dd_{\ad_1}\D_{\a_1}\S\right.\\
&+B^{\a(k)\ad(k+1)}_{l}~\Box^{l}~\Dd_{\ad_{k+1}}\D_{\a_k}\Dd_{\ad_k}\dots\D_{\a_1}\Dd_{\ad_1}\S 
\n\label{tr1} \\
&+\Gamma^{\a(k)\ad(k)}_{l}~\Box^{l}~\Dd_{\ad_k}\D_{\a_k}\dots\Dd_{\ad_1}\D_{\a_1}\S\\
&+\left.\vphantom{\frac12}\Delta^{\a(k)\ad(k)}_{l}~\Box^{l}~\D_{\a_k}\Dd_{\ad_k}\dots\D_{\a_1}\Dd_{\ad_1}\S\right\}
\eea
which depends on four infinite families of parameters 
$\{A_{\a(k+1)\ad(k)}^{l},~B_{\a(k)\ad(k+1)}^{l},~\Gamma_{\a(k)\ad(k)}^{l},~\Delta_{\a(k)\ad(k)}^{l}~\}$
with independently symmetrized dotted and undotted indices. 
To have this transformation consistent with the linearity of $\S$ 
($\Dd^2\S=0$), we must have
\bea{l}\n\label{LinearCon}
A^{l}_{\a(k+1)\ad(k)}=-\Dd^{\ad_{k+1}}\Gamma^{l}_{\a(k+1)\ad(k+1)}+\tfrac{1}{k+2}\Dd^{\ad_{k+1}}\Delta^{l}_{\a(k+1)\ad(k+1)}~,~k=0,1...~,\sn\vspace{1ex}\\
\tfrac{1}{(k+1)!}\Dd_{(\ad_{k+1}}\Delta_{\a(k)\ad(k))}^{l}=-\Dd^2 
B_{\a(k)\a(k+1)}^{l}~,~k=1,2,...~,\sn\vspace{1ex}\\
\Dd_{\ad}(\Gamma^{l}+\Delta^{l})=-\Dd^2 B^{l}_{\ad}~.\sn
\eea
It is worth mentioning that (\ref{tr1}) is not the most general transformation 
linear in $\S$ 
one can write.
For example we can have $\D^2$, $\Dd^2\D^2$ or $\Dd^2\D_{\a_{k+1}}$ 
appropriately placed in the various terms.
These extra terms will modify the above constraints, nevertheless we have verified that they do not
introduce extra structure regarding the coupling to higher spin supermultiplets, so 
we will not consider them. Solving (\ref{LinearCon}) we conclude that the parameters 
$B^{l}_{\a(k)\ad(k+1)},~\Gamma^{l}_{\a(k)\ad(k)}$ are unconstrained, whereas 
$A^{l}_{\a(k+1)\ad(k)},~\Delta^{l}_{\a(k)\ad(k)}$
are given by
\bea{l}\n\label{LinConSol}
A^{l}_{\a(k+1)\ad(k)}=-\Dd^{\ad_{k+1}}\Gamma^{l}_{\a(k+1)\ad(k+1)}+\tfrac{1}{k+1}\Dd^2\ell^{l}_{\a(k+1)\ad(k)}~,~k=0,1...~,\sn\vspace{1ex}\\
\Delta^{l}_{\a(k)\ad(k)}=\Dd^{\ad_{k+1}}B^{l}_{\a(k)\ad(k+1)}+\tfrac{1}{k!}\Dd_{(\ad_k}\ell^{l}_{\a(k)\ad(k-1))}~,~k=1,2,..~,\sn\vspace{1ex}\\
\Gamma^{l}+\Delta^{l}=\Dd^{\ad}B^{l}_{\ad}+\Dd^2\ell^{l}~.\sn
\eea
where $\ell^{l}_{\a(k+1)\ad(k)}$ and $\ell^{l}$ are arbitrary, unconstrained superfields.
A useful observation is that the constraints (\ref{LinearCon}) do not mix the 
different $l$-levels and all the determined parameters (\ref{LinConSol})
are functions of other parameters of the same level. Therefore, the $l$ label 
does not provide any further structure and we can simplify (\ref{tr1}) by 
confining ourselves in the $l=0$ level\footnote{From this point forward we drop 
the $l$ label.}. Therefore, the transformation we consider is
\bea{ll}
\d_{1}\S=g\sum_{k=0}^{\infty}&\left\{\vphantom{\frac12}
~~\Dd^2\ell^{\a(k+1)\ad(k)}~\tfrac{i^k}{k+1}~\D_{\a_{k+1}}\pa_{\a_k 
\ad_k}\dots\pa_{\a_1 \ad_1}\S\right.\\
&~+\tfrac{1}{(k+1)!}~\Dd^{(\ad_{k+1}}\ell^{\a(k+1)\ad(k))}~i^k~\D_{\a_{k+1}}\Dd_{\ad_{k+1}}\pa_{\a_k 
\ad_k}\dots\pa_{\a_1 \ad_1}\S\\
&~+B^{\a(k)\ad(k+1)}~i^k~\Dd_{\ad_{k+1}}\pa_{\a_k \ad_k}\dots\pa_{\a_1 \ad_1}\S 
\n\label{tr1a} \\
&~-\Dd_{\ad_{k+2}}B^{\a(k+1)\ad(k+2)}~i^k~\D_{\a_{k+1}}\Dd_{\ad_{k+1}}\pa_{\a_k 
\ad_k}\dots\pa_{\a_1 \ad_1}\S\\
&~+\Gamma^{\a(k+1)\ad(k+1)}~i^k~\Dd_{\ad_{k+1}}\D_{\a_{k+1}}\pa_{\a_k 
\ad_k}\dots\pa_{\a_1 \ad_1}\S\\
&~+\left.\vphantom{\frac12}\Dd_{\ad_{k+1}}\Gamma^{\a(k+1)\ad(k+1)}~i^k~\D_{\a_{k+1}}\pa_{\a_k 
\ad_k}\dots\pa_{\a_1 \ad_1}\S\right\}\\
&\hspace{-6ex}+g~\left(\Dd^2\ell - \Dd_{\ad}B^{\ad}\right)~\S~.
\eea
Additionally, if we want to make contact with the superdiffeomorphism transformation,
we choose
\bea{l}\n\label{FixG}
\Gamma_{\a(k)\a(k)}=\Delta_{\a(k)\ad(k)}~,~k=1,2,... ~.
\eea
so that the $\Gamma_{\a(k)\ad(k)}$ with $\Delta_{\a(k)\ad(k)}$ terms for $k\geq1$ combine to give
spacetime derivatives resulting in
\bea{ll}
\d_{1}\S=-g\sum_{k=0}^{\infty}&\left\{\vphantom{\frac12}
~~~\Dd^2\ell^{\a(k+1)\ad(k)}~i^{k}~\D_{\a_{k+1}}\pa_{\a_{k}\ad_{k}}\dots\pa_{\a_1\ad_1}\S\right.\\
&~~-\tfrac{1}{(k+1)!}\Dd^{(\ad_{k+1}}\ell^{\a(k+1)\ad(k))}~i^{k+1}~\pa_{\a_{k+1}\ad_{k+1}}\dots\pa_{\a_1\ad_1}\S\\
\n\label{tr2} \\
&~~-B^{\a(k)\ad(k+1)}~i^{k}~\Dd_{\ad_{k+1}}\pa_{\a_{k}\ad_{k}}\dots\pa_{\a_1\ad_1}\S\\
&~~+\left.\vphantom{\frac12}\Dd_{\ad_{k+2}}B^{\a(k+1)\ad(k+2)}~i^{k+1}~\pa_{\a_{k+1}\ad_{k+1}}\dots\pa_{\a_1\ad_1}\S\right\}\\
&\hspace{-6ex}+g(\Dd^2\ell-\Dd_{\ad}B^{\ad})~\S~~.
\eea

%%%%%%%%%%%%%%%%%%%%%%%%%%%%%%%%%%%%%%%%%%%%%%
%%%%%%%%%%%%%%%%%%  %%%%%%%%%%%%%%%%%%%%%%%%%%
\section{Higher Spin supercurrent multiplet}\label{CM}
Having found the appropriate first order transformation for the complex linear 
superfield,
we use it to perform Noether's procedure for the cubic order terms, as described
in \textsection\ref{dSigma} and construct the higher spin supercurrent
multiplet. Starting from the free theory (\ref{freeth}) we calculate its 
variation under (\ref{tr2}):
\bea{l}
\d_{1} S_{0}=g\int d^8z\sum_{k=0}^{\infty}\left\{\vphantom{\frac12}
~\Dd^2\ell^{\a(k+1)\ad(k)}~i^{k}~\D_{\a_{k+1}}\pa_{\a_{k}\ad_{k}}\dots\pa_{\a_1\ad_1}\S~\Sd~+~c.c.\right.\\
\hspace{21ex}+\bar{B}^{\a(k+1)\ad(k)}~(-i)^{k}~\D_{\a_{k+1}}\S~\pa_{\a_{k}\ad_{k}}\dots\pa_{\a_1\ad_1}\Sd~+~c.c.\n\label{dS1}\vspace{1ex}\\
\hspace{21ex}-\left.\vphantom{\frac12}\tfrac{1}{(k+1)!}\Dd^{(\ad_{k+1}}\ell^{\a(k+1)\ad(k))}~i^{k+1}~
\pa_{\a_{k+1}\ad_{k+1}}\dots\pa_{\a_1\ad_1}\S~\Sd~+~c.c.\right\}\vspace{1ex}\\
\hspace{7ex}-g\int d^8z~(\Dd^2\ell+\D^2\bar{\ell})~\S~\Sd~~.
\eea
At this point it is tempting to choose the $B_{\a(k)\ad(k+1)}$ parameter to be a 
function of the $\ell_{\a(k+1)\ad(k)}$ parameter. In principle we can write all 
possible terms allowed by engineering dimensions and index structure:
\bea{l}\n
\bar{B}_{\a(k+1)\ad(k)}=d_1\Dd^2\ell_{\a(k+1)\ad(k)}+\tfrac{d_2}{(k+1)!}\D_{(\a_{k+1}}\Dd^{\ad_{k+1}}\bar{\ell}_{\a(k))\ad(k+1)}
+\tfrac{d_3}{(k+1)!}\Dd^{\ad_{k+1}}\D_{(\a_{k+1}}\bar{\ell}_{\a(k))\ad(k+1)}~.
\eea
A similar step was done in \cite{cls} and led to the coupling of the complex 
linear theory to the various formulations of supergravity depending on the 
values of $d_1,~d_2,~d_3$. In this analysis supergravity would correspond to 
$k=0$. Notice that in (\ref{dS1}) the coefficients of $\bar{B}_{\a(k+1)\ad(k)}$ 
and $\Dd^2\ell_{\a(k+1)\ad(k)}$ do not match for $k\geq1$, hence we can not 
repeat the same arguments in \cite{cls} for higher spin coupling. There is only 
one viable option
\bea{l}
\bar{B}_{\a(k+1)\ad(k)}=-\Dd^2\ell_{\a(k+1)\ad(k)}\n\label{fixB}~.
\eea
By doing this, the variation of the action can be written in the following way
\bea{l}
\d_{1} S_{0}=g\int d^8z\sum_{k=0}^{\infty}\left\{\vphantom{\frac12}
~\Dd^2\ell^{\a(k+1)\ad(k)}~i^{k}\left[\pa^{(k)}\D\S~\Sd ~+(-1)^{k-1}~
\D\S~\pa^{(k)}\Sd\right]+c.c.\right.\vspace{1ex}\\
\hspace{21ex}-\left.\vphantom{\frac12}\tfrac{1}{(k+1)!}\Dd^{(\ad_{k+1}}\ell^{\a(k+1)\ad(k))}~
i^{k+1}~\pa^{(k+1)}\S~\Sd~+~c.c.\right\}\n\label{dS2}\vspace{1ex}\\
\hspace{7ex}-g\int d^8z~(\Dd^2\ell+\D^2\bar{\ell})~\S~\Sd~~
\eea
where for simplicity we suppress the uncontracted indices, their
symmetrization and the symmetrization factors when they are appropriate.
The symbol $\pa^{(k)}$ denotes a string of $k$ spacetime derivatives. Moreover, 
we can use the following identity
\bea{l}
\pa^{(k)}A~B~+(-1)^{k-1}~A~\pa^{(k)}B=\pa\left\{\sum_{n=0}^{k-1}(-1)^{n}~\pa^{(k-1-n)}A~\pa^{(n)}B\right\}\n\label{identity}
\eea
which holds for arbitrary (super)functions $A$ and $B$ and simplify (\ref{dS2}) 
to:
\bea{l}
\d_{1} S_{0}=-g\int d^8z\sum_{k=0}^{\infty}\left\{\vphantom{\frac12}
~\Dd^2\ell^{\a(k+1)\ad(k)}~\T_{\a(k+1)\ad(k)}+c.c.\right.\vspace{1ex}\\
\hspace{23ex}-\left.\vphantom{\frac12}\tfrac{1}{(k+1)!}\Dd^{(\ad_{k+1}}\ell^{\a(k+1)\ad(k))}~
\J_{\a(k+1)\ad(k+1)}~+~c.c.\right\}\n\label{dS3}\vspace{1ex}\\
\hspace{7ex}+g\int d^8z~(\Dd^2\ell+\D^2\bar{\ell})~\J~~.
\eea
where
\bea{l}\n
\J_{\a(k+1)\ad(k+1)}=-i^{(k+1)}~\pa^{(k+1)}\S~\Sd
+\tfrac{1}{(k+1)!}\D_{(\a_{k+1}}\Dd^2\bar{U}_{\a(k))\ad(k+1)}+\tfrac{k+1}{(k+2)!}\Dd_{(\ad_{k+1}}W_{\a(k+1)\ad(k))}~,~\sn\label{J}\\
\T_{\a(k+1)\ad(k)}=-i^{(k-1)}~\D\Dd\left\{\sum_{n=0}^{k-1}(-1)^{n}~\pa^{(k-1-n)}\D\S~\pa^{(n)}\Sd\right\}
+\tfrac{1}{(k+1)!}\D_{(\a_{k+1}}\Dd^{\ad_{k+1}}\bar{U}_{\a(k))\ad(k+1)}\sn\label{T}\\
\hspace{13ex}+W_{\a(k+1)\ad(k)}~,\\
\J=-\S\Sd~.\sn
\eea
The superfields $W_{\a(k+1)\ad(k)}$ and $\bar{U}_{\a(k)\ad(k+1)}$ are 
improvement terms that we can add, as discussed in \cite{hsch1}.
Also, it is important to keep in mind that the 
$\J_{\a(k+1)\ad(k+1)},~\T_{\a(k+1)\ad(k)}$ and $\J$ are not defined uniquely but 
they satisfy an equivalence relation. For example
\bea{l}
\T_{\a(k+1)\ad(k)}\sim \T_{\a(k+1)\ad(k)}+ 
\Dd_{(\ad_{k}}P^{(1)}_{\a(k+1)\ad(k-1))}+\Dd^{\ad_{k+1}}P^{(2)}_{\a(k+1)\ad(k+1)}\n\label{eqrelT}\\
\hspace{23.6ex}+\D_{(\a_{k+1}}\Dd^2P^{(3)}_{\a(k))\ad(k)}+\D^{\a_{k+2}}\Dd^2 
P^{(4)}_{\a(k+2)\ad(k)}~.
\eea
for arbitrary superfields $P^{(1)},~P^{(2)},~P^{(3)},~P^{(4)}$.\\
Using (\ref{identity}) we can express the $-i^{k+1}\pa^{k+1}\S~\Sd$ in the 
following manner
\bea{l}
-i^{(k+1)}~\pa^{k+1}\S~\Sd = 
X_{\a(k+1)\ad(k+1)}+\tfrac{1}{(k+1)!^2}\Dd_{(\ad_{k+1}}\D_{(\a_{k+1}}Z_{\a(k))\ad(k))}\sn\vspace{1ex}\n\label{decompXZ}
\eea
where $X_{\a(k+1)\ad(k+1)}$ and $Z_{\a(k)\ad(k)}$ are real superfields given 
by the following expressions:
\bea{l}\n
X_{\a(k+1)\ad(k+1)}=-\tfrac{i^{k+1}}{2}\left[\vphantom{\frac12}~\pa^{(k+1)}\S~\Sd~+(-1)^{k+1}~\S~\pa^{(k+1)}\Sd~\right]\sn\label{X}\\
\hspace{16ex}-\tfrac{i^{k}}{2}\sum_{n=0}^{k}(-1)^{n}\left\{\vphantom{\frac12}~\pa^{(k-n)}[\D,\Dd]\S~\pa^{(n)}\Sd~
+~\pa^{(k-n)}\S~\pa^{(n)}[\D,\Dd]\Sd~\right\}\\
\hspace{16ex}-i^{k}\sum_{n=0}^{k}(-1)^{n}\left\{\vphantom{\frac12}~\pa^{(k-n)}\D\S~\pa^{(n)}\Dd\Sd~-~\pa^{(k-n)}\Dd\S~\pa^{(n)}\D\Sd~\right\}~~,\\
Z_{\a(k)\ad(k)}=-i^{k}\sum_{n=0}^{k}(-1)^{n}~\pa^{(k-n)}\S~\pa^{(n)}\Sd\sn\label{Z}~~.
\eea
Because of (\ref{decompXZ}), it is straight forward to prove that 
there is always a choice for the improvement term $W_{\a(k+1)\ad(k)}$
that makes $\J_{\a(k+1)\ad(k)}$ real. Specifically, by 
selecting
\bea{l}
W_{\a(k+1)\ad(k)}=-\tfrac{k+2}{k+1}\D^2U_{\a(k+1)\ad(k)}-\tfrac{k+2}{k+1}\tfrac{1}{(k+1)!}\D_{(\a_{k+1}}Z_{\a(k))\ad(k)}\n
\eea
we get
\bea{l}\n
\J_{\a(k+1)\ad(k+1)}=X_{\a(k+1)\ad(k)}
+\tfrac{1}{(k+1)!}\D_{(\a_{k+1}}\Dd^2\bar{U}_{\a(k))\ad(k+1)}-\tfrac{1}{(k+1)!}\Dd_{(\ad_{k+1}}\D^2 
U_{\a(k+1)\ad(k))}~,~\sn\label{supercurrent}\vspace{1ex}\\
\T_{\a(k+1)\ad(k)}=\tfrac{1}{(k+1)!}\D_{(\a_{k+1}}\T_{\a(k))\ad(k)}\sn\label{tsd}~,~\vspace{1ex}\\
\T_{\a(k)\ad(k)}=-i^{(k-1)}~\Dd\left\{\sum_{n=0}^{k-1}(-1)^{n}~\pa^{(k-1-n)}\D\S~\pa^{(n)}\Sd\right\}
-\tfrac{k+2}{k+1}Z_{\a(k)\ad(k)}\sn\label{supertrace}\\
\hspace{11ex}+\tfrac{k+2}{k+1}~\D^{\a_{k+1}}U_{\a(k+1)\ad(k)}+\Dd^{\ad_{k+1}}\bar{U}_{\a(k))\ad(k+1)}~.
\eea
The reality of $\J_{\a(k+1)\ad(k+1)}$ and the fact that $\T_{\a(k+1)\ad(k)}$ 
becomes a total spinorial derivative (\ref{tsd}), allows us to modify 
(\ref{dS3}) in the following way:
\bea{l}
\d_{1} S_{0}=-g\int d^8z\sum_{k=0}^{\infty}\left\{\vphantom{\frac12}
~~~~\left[\Dd^2\ell^{\a(k+1)\ad(k)}-\D_{\a_{k+2}}\lambda^{\a(k+2)\ad(k)}\right]
~\tfrac{1}{(k+1)!}\D_{(\a_{k+1}}\T_{\a(k))\ad(k)}+c.c.\right.\vspace{1ex}\\
\hspace{23ex}+\left.\vphantom{\frac12}\left[\tfrac{1}{(k+1)!}\D^{(\a_{k+1}}\bar{\ell}^{\a(k))\ad(k+1)}
-\tfrac{1}{(k+1)!}\Dd^{(\ad_{k+1}}\ell^{\a(k+1)\ad(k))}\right]~\J_{\a(k+1)\ad(k+1)}~\right\}\n\label{dS4}\vspace{1ex}\\
\hspace{7ex}+g\int d^8z~(\Dd^2\ell+\D^2\bar{\ell})~\J~~.
\eea
The terms inside the square brackets are exactly the zeroth order 
transformations that appear in (\ref{hstr2}).
Hence, in order to get an invariant theory, we must add the following cubic 
interaction terms between $(s+1,s+1/2)$ higher spin supermultiplets 
 (\ref{hstr2}) and the complex linear superfield
\bea{l}
S_{\text{HS - $\S$ cubic interactions}}=g\int 
d^8z~\sum_{k=0}^{\infty}\left\{\vphantom{\frac12}~~~H^{\a(k+1)\ad(k+1)}\J_{\a(k+1)\ad(k+1)}\right.\n
\label{cubint}\\
\hspace{37ex}+\left.\vphantom{\frac12}\chi^{\a(k+1)\ad(k)}\D_{\a_{k+1}}\T_{\a(k)\ad(k)}
+\bar{\chi}^{\a(k)\ad(k+1)}\Dd_{\ad_{k+1}}\bar{\T}_{\a(k)\ad(k)}\right\}\\
\hspace{21ex}-g\int d^8z~V\J
\eea
where $V$ ($\d_{0}V=\Dd^2\ell+\D^2\bar{\ell}$) is the real prepotential used for 
the description of the vector supermultiplet $(1,1/2)$.

Superfields $\J_{\a(k+1)\ad(k+1)}$ (\ref{supercurrent}) and $\T_{\a(k)\ad(k)}$ 
(\ref{supertrace}) are the $\S$-generated 
higher spin supercurrent and higher spin supertrace respectively and together 
they form the higher spin supercurrent multiplet
$\left\{\J_{\a(k+1)\ad(k+1)}~,~\T_{\a(k)\ad(k)}\right\}$. The results we get are 
similar to the case of the chiral superfield 
described in \cite{hsch1} and \cite{hsch2}. 
Specifically, we find that a single, free, massless, complex linear supermultiplet 
couples only to half integer superspin supermultiplets with a preference to the 
transverse formulation (\ref{hstr2}) which is the only formulation of 
half-integer superspins that can be elevated to $\N=2$ theories. Furthermore, 
like in the chiral case the higher spin supercurrent and supertrace include 
higher derivative terms, as expected from \cite{Metsaev}.

%%%%%%%%%%%%%%%%%%%%%%%%%%%%%%%%%%%%%%%%%%%%%%
%%%%%%%%%%%%%%%%%%  %%%%%%%%%%%%%%%%%%%%%%%%%%
\section{Minimal higher spin supercurrent multiplet}\label{MMM}
In the previous section we constructed what is known as the \emph{canonical}
supercurrent multiplet \cite{Magro}. In this section we will prove that for 
every value of the non-negative integer
$k$ there is a unique, alternative higher spin supercurrent multiplet, 
called the \emph{minimal} supercurrent multiplet
defined by $\T^{\textit{min}}_{\a(k)\ad(k)}=0$. This will be possible due to 
(\emph{i}) the freedom in choosing
the unconstrained improvement term $U_{\a(k+1)\ad(k)}$, (\emph{ii}) the freedom 
in the definition of $\T_{\a(k+1)\ad(k)}$ (\ref{eqrelT})
and (\emph{iii}) the freedom to absorb trivial terms by redefining $\S$.

For an arbitrary superfield $\Y_{\ad}$, the combination $\Dd^{\ad}\Y_{\ad}$ is a 
complex linear superfield.
Therefore, one can consider the redefinition of $\S$
\bea{l}
\S=\hat{\S}+g~\Dd^{\ad}\Y_{\ad}\n\label{redefSigma}
\eea
which will modify the free theory in the following manner
\bea{l}
S_{0}=-\int d^8z ~\hat{\S}~\bar{\hat{\S}}~+g\int d^8z 
~\bar{\Y}^{\a}~\D_{\a}\hat{\S}~+g\int d^8z ~\Y^{\ad}~\Dd_{\ad}\bar{\hat{\S}}
~+g^2\int d^8z ~\Y^{\ad}~\Dd_{\ad}\D^{\a}\bar{\Y}_{\a}\n\label{redefAction}
\eea
where $g$ is the perturbative parameter. Because we are working up to order 
$g^1$ (cubic interactions) we can ignore the last term in (\ref{redefAction}) 
and therefore conclude that terms in the action that depend on $\D\S$ or 
$\Dd\Sd$ (these are the on-shell equation of motion for the free theory)
are \emph{trivial} terms and can be absorbed by an appropriate redefinition of 
$\S$ like (\ref{redefSigma}). This realization simplifies the calculation of the 
supercurrent multiplet. For example, if we consider the transformation 
of $\Sigma$ (\ref{tr1a}) before fixing $\Gamma_{\a(k)\ad(k)}$ and 
$B_{\a(k)\ad(k+1)}$ then we get:
\bea{l}
\d_{1} S_{0}=-g\int d^8z\sum_{k=0}^{\infty}\left\{\vphantom{\frac12}
~\Dd^2\ell^{\a(k+1)\ad(k)}~\tfrac{i^k}{k+1}~\pa^{(k)}\D\S~\Sd\right.\\
\hspace{22ex}+\tfrac{1}{(k+1)!}\Dd^{(\ad_{k+1}}\ell^{\a(k+1)\ad(k))}~i^{(k+1)}~\pa^{(k+1)}\S~\Sd\\
\hspace{22ex}-\tfrac{1}{(k+1)!}\Dd^{(\ad_{k+1}}\ell^{\a(k+1)\ad(k))}~i^k~\pa^{(k)}\Dd\D\S~\Sd\n\\
\hspace{22ex}-B^{\a(k)\ad(k+1)}~i^k~\pa^{(k)}\S~\Dd\Sd\\
\hspace{22ex}+B^{\a(k+1)\ad(k+2)}~i^{k}~\pa^{(k)}\Dd\D\S~\Dd\Sd\\
\hspace{22ex}+\left.\vphantom{\frac12}\Gamma^{\a(k+1)\ad(k+1)}~i^k~\pa^{(k)}\D\S~\Dd\Sd\right\}~+c.c.\\
\hspace{7ex}-g\int d^8z~ (\Dd^2\ell+\D^2\bar{\ell})~\S~\Sd~.
\eea
Observe that the contribution of the $B_{\a(k)\ad(k+1)}$ and $\Gamma_{\a(k)\ad(k)}$ 
terms is trivial in the sense explained previously, hence we can remove them by
an appropriate redefinition of $\S$. This is very satisfying because, unlike the 
chiral case, the first order transformation of $\S$ (\ref{tr1a}) is far from 
unique. It includes two infinite families of unconstrained superfields. Despite 
that, as we see at the level of the action, these unconstrained parameters
do not give non-trivial contributions. Therefore, the variation of
the action is completely fixed without any freedom left:
\bea{l}
\d_{1} S_{0}=-g\int d^8z\sum_{k=0}^{\infty}
\left[-\tfrac{1}{(k+1)!}\Dd^{(\ad_{k+1}}\ell^{\a(k+1)\ad(k))}\right]~\J^{\textit{min}}_{\a(k+1)\ad(k+1)}~+~c.c.\n\label{dSmin}\\
\hspace{7ex}+g\int d^8z~ (\Dd^2\ell +\D^2\bar{\ell})~\J
\eea
where
\bea{l}\n
\J^{\textit{min}}_{\a(k+1)\ad(k+1)}=-i^{(k+1)}~\pa^{(k+1)}\S~\Sd\sn\label{Jmin1}\\
\hspace{15.5ex}+\tfrac{1}{(k+1)!}\D_{(\a_{k+1}}\Dd^2\bar{U}_{\a(k))\ad(k+1)}
-\tfrac{k+1}{(k+2)!(k+1)!)}\Dd_{(\ad_{k+1}}\D_{(\a_{k+1}}\Dd^{\bd}\bar{U}_{\a(k))\bd\ad(k))}~,~\\
\J=-\S~\Sd\sn~.
\eea
The superfield $U_{\a(k+1)\ad(k)}$ is an improvement term similar to the improvement terms 
appearing in (\ref{J}). We can deduce its presence
by constraining  the $W_{\a(k+1)\ad(k)}$ superfield in 
(\ref{T}) to cancel the $U_{\a(k+1)\ad(k)}$ contribution, giving 
$W_{\a(k+1)\ad(k)}=-\tfrac{1}{(k+1)!}\D_{(\a_{k+1}}\Dd^{\ad_{k+1}}\bar{U}_{\a(k))\ad(k+1)}$.
Notice, there is no $\T_{\a(k+1)\ad(k)}$ contribution like in (\ref{dS3}) which 
means we are already in the \emph{minimal} setup of the supercurrent multiplet. 
Also, we must keep in mind that the above definition of 
$\J^{\textit{min}}_{\a(k+1)\ad(k+1)}$ satisfies an equivalence relation
\bea{l}
\J^{\textit{min}}_{\a(k+1)\ad(k+1)}\sim\J^{\textit{min}}_{\a(k+1)\ad(k+1)}
+\Dd^{\ad_{k+2}}\Theta^{(1)}_{\a(k+1)\ad(k+2)}+\Dd^2\Theta^{(2)}_{\a(k+1)\ad(k+1)}\n~.
\eea
for arbitrary $\Theta^{(1)}$ and $\Theta^{(2)}$ superfields.
Furthermore, using (\ref{decompXZ}) we can rewrite (\ref{Jmin1})
\bea{l}
\J^{\textit{min}}_{\a(k+1)\ad(k+1)}=X_{\a(k+1)\ad(k+1)}+\tfrac{1}{(k+1)!}\D_{(\a_{k+1}}\Dd^2\bar{U}_{\a(k))\ad(k+1)}
-\tfrac{1}{(k+1)!}\Dd_{(\ad_{k+1}}\D^2U_{\a(k))\ad(k+1)}\n\label{Jmin2}\\
\hspace{14ex}+\tfrac{1}{(k+1)!^2}\Dd_{(\ad_{k+1}}\D_{(\a_{k+1}}\left[Z_{\a(k)\ad(k)}-\tfrac{k+1}{k+2}\Dd^{\bd}\bar{U}_{\a(k))\bd\ad(k))}
-\D^{\b}U_{\b\a(k))\ad(k))}\right]~.
\eea
Thus, we can make $\J^{\textit{min}}_{\a(k+1)\ad(k+1)}$ real
if and only if we can find a superfield $U_{\a(k+1)\ad(k)}$ that satisfies the 
following constraint:
\bea{l}
Z_{\a(k)\ad(k)}=\D^{\a_{k+1}}U_{\a(k+1)\ad(k)}+\tfrac{k+1}{k+2}\Dd^{\ad_{k+1}}\bar{U}_{\a(k)\ad(k+1)}\n\label{UCond}\\
\hspace{10ex}+\tfrac{1}{k!}\D_{(\a_{k}}\zeta^{(1)}_{\a(k-1))\ad(k)}+\D^2\zeta^{(2)}_{\a(k)\ad(k)}
+\Dd^2\zeta^{(3)}_{\a(k)\ad(k)}+\zeta^{(4)}_{\a(k)\ad(k)}(\D\S, 
\Dd\Sd)
\eea
for some superfields $\zeta^{(1)}_{\a(k-1)\ad(k)}$, $\zeta^{(2)}_{\a(k)\ad(k)}$, 
$\zeta^{(3)}_{\a(k)\ad(k)}$ and a term $\zeta^{(4)}_{\a(k)\ad(k)}$ that 
depends only on $\D\S$ and $\Dd\Sd$. This condition corresponds to setting the 
supertrace (\ref{supertrace}) to zero, which is exactly the demand for a minimal 
supercurrent multiplet.

Consider the following ansatz for a solution of 
(\ref{UCond})
\bea{l}
U_{\a(k+1)\ad(k)}=\sum_{p=0}^{k}~f_{p}~\pa^{(p)}\S~\pa^{k-p}\lambda\n\label{Uansatz}
\eea
with $\bar{\lambda}_{\ad}$
being the unconstrained prepotential of the complex linear superfield $\S$ 
($\S=\Dd^{\ad}\bar{\lambda}_{\ad}$, $\Sd=\D^{\a}\lambda_{\a}$).
It is straight forward to show that\footnote{We remind the reader of our convention to suppress 
uncontracted indices, their symmetrization and appropriate factors.}:
\bea{l}
\D^{\a_{k+1}}U_{\a(k+1)\ad(k)}=~~\sum_{p=0}^{k}~f_{p}~\tfrac{k-p+1}{k+1}~\pa^{p}\S~\pa^{(k-p)}\Sd\n\\
\hspace{18ex}-i\sum_{p=0}^{k-1}~f_{p}~\tfrac{p+1}{k+1}~\pa^{p}\Dd\S~\pa^{(k-p)}\D\Sd~
+\D^2\left[\dots\right]+\mathcal{O}(\D\S , \Dd\Sd)
\eea
and 
\bea{l}
\D^{\a_{k+1}}U_{\a(k+1)\ad(k)}+\tfrac{k+1}{k+2}~\Dd^{\ad_{k+1}}\bar{U}_{\a(k)\ad(k+1)}=
~~\left\{\vphantom{\frac12}~f_{0}+f_{k}^{*}\tfrac{1}{k+2}~\right\}~\S~\pa^{(k)}\Sd\n\\
\hspace{42ex}+\sum_{p=1}^{k}\left\{\vphantom{\frac12}~f_{p}+f_{k-p}^{*}\tfrac{p+1}{k+2}-f_{k-p+1}^{*}\tfrac{k-p+1}{k+2}~\right\}
~\pa^{(p)}\S~\pa^{(k-p)}\Sd\\
\hspace{42ex}+\tfrac{1}{k!}\D_{(\a_k}\left[\dots\right]+\D^2\left[\dots\right]+\Dd^2\left[\dots\right]
+\mathcal{O}(\D\S , \Dd\Sd)~.
\eea
From the above and equation (\ref{Z}), we get that condition (\ref{UCond}) takes 
the form
\bea{l}\n\label{f}
(k+2)f_{0}+f_{k}^{*}=(-1)^{k+1}~i^{k}~(k+2)\sn~,~\\
(k+2)f_{p}+(p+1)f_{k-p}^{*}-(k-p+1)f_{k-p+1}^{*}=(-1)^{k-p+1}~i^{k}~(k+2)~,~\sn\\
\hspace{60ex}p=1,\dots,k~.
\eea
This is a system of $k+1$ linear equations for the $k+1$ complex variables 
$f_{p}, ~p=0,1,\dots,k$. If we introduce a new set of variables $\hat{f}_{p}$ 
defined as
\bea{l}
f_{p}=(-1)^{k+p+1}~i^{k}~+~(-1)^{k}~\hat{f}_{k-p}\n
\eea
then the (\ref{f}) system of equations takes the form
\bea{l}\n\label{fhat}
(k+2)\hat{f}_{k}+\hat{f}_{0}^{*}=i^{k}\sn~,~\\
(k+2)\hat{f}_{p}+(k-p+1)\hat{f}_{k-p}^{*}-(p+1)\hat{f}_{k-p-1}^{*}=(-1)^{k+p}~i^{k}~(k+2)~,~\sn\\
\hspace{56ex}p=0,\dots,k-1~.
\eea
This is the same system of equations which appeared in \cite{hsch1} in 
the process of removing the supertrace. There it was proved that for 
every $k$, this system of equations has a solution, hence we conclude that the condition 
(\ref{UCond}) can always be satisfied. Using the solution coming from \cite{hsch1},
we can write
\bea{l}
f_{p}=(-1)^{k+p+1}~i^{k}~\left[\vphantom{\frac12}~1~-~\frac{\sum\limits_{j=0}^{p}\binom{k+j+1}{p}\binom{k+1-j}{p-j}}{\binom{2k+3}{k+2}}\right]~~,~~
p=0,1,\dots,k~.\n
\eea
Therefore, due to (\ref{dSmin}) with a real 
$\J^{\textit{min}}_{\a(k+1)\ad(k+1)}$ we must add the following interaction 
terms
\bea{l}
S_{\text{HS-$\Sigma$ ~minimal cubic interactions}}=g\int 
d^8z~\sum_{k=0}^{\infty}~H^{\a(k+1)\ad(k+1)}\J^{\textit{min}}_{\a(k+1)\ad(k+1)}~-~g\int 
d^8z~ V\J\n\label{mincubint}~.
\eea

The conclusion is that, similar to the chiral case, for any value of $k$, we can 
go from the \emph{canonical} $\{\J_{\a(k+1)\ad(k+1)}~,~\T_{\a(k)\ad(k)}\}$ 
(\ref{supercurrent}, \ref{supertrace})
to the \emph{minimal} higher spin supercurrent multiplet 
$\{\J^{\textit{min}}_{\a(k+1)\ad(k+1)}~,~0\}$. To get the explicit expression 
for
$\J^{\textit{min}}_{\a(k+1)\ad(k+1)}$ we combine (\ref{X}, \ref{supercurrent}, 
\ref{Uansatz})
\bea{l}
\J^{\textit{min}}_{\a(k+1)\ad(k+1)}=-i\left\{f_{0}^{*}+i^{k}\right\}~\pa^{(k+1)}\S~\Sd~+~i\left\{f_{0}+(-i)^{k}\right\}~\S~\pa^{(k+1)}\Sd\\
\hspace{16ex}+i\sum_{p=1}^{k}\left\{f_{p}-f_{k+1-p}^{*}+(-1)^{k+p}~i^k\right\}~\pa^{(p)}\S~\pa^{(k+1-p)}\Sd\n\label{JminF}\\
\hspace{16ex}+\sum_{p=0}^{k}\left\{f_{p}+f_{k-p}^{*}+(-1)^{k+p}~i^k\right\}~\pa^{(p)}\Dd\S~\pa^{(k-p)}\D\Sd~.
\eea
The first three ($k=0,~k=1,~k=2$) minimal supercurrents are:
\bea{l}
\J^{\textit{min}}_{\a\ad}=-\tfrac{i}{3}~\pa_{\a\ad}\S~\Sd+\tfrac{i}{3}~\S~\pa_{\a\ad}\Sd-\tfrac{1}{3}~\Dd\S~\D\Sd\n\label{k0}~,~\vspace{1ex}\\
\J^{\textit{min}}_{\a\b\ad\bd}=\tfrac{1}{10}~\pa^{(2)}\S~\Sd+\tfrac{1}{10}~\S~\pa^{(2)}\Sd-\tfrac{2}{5}~\pa\S~\pa\Sd
+\tfrac{i}{5}~\Dd\S~\pa\D\Sd-\tfrac{i}{5}~\pa\Dd\S~\D\Sd\n\label{k1}~,~\vspace{1ex}\\
\J^{\textit{min}}_{\a\b\g\ad\bd\gd}=\tfrac{i}{35}~\pa^{(3)}\S~\Sd-\tfrac{i}{35}~\S~\pa^{(3)}\Sd+\tfrac{9i}{35}~\pa\S~\pa^{(2)}\Sd
-\tfrac{9i}{35}~\pa^{(2)}\S~\pa\Sd\n\label{k2}\\
\hspace{11ex}+\tfrac{3}{35}~\Dd\S~\pa^{(2)}\D\Sd-\tfrac{9}{35}~\pa\Dd\S~\pa\D\Sd+\tfrac{3}{35}~\pa^{(2)}\Dd\S~\D\Sd~.
\eea
The above expressions (\ref{JminF}, \ref{k0}, \ref{k1}, \ref{k2}) for the the 
minimal higher spin supercurrent of a complex linear superfield have striking 
similarities with the minimal higher spin supercurrent for a chiral superfield. 
The detailed connection will be established in \textsection\ref{Duality}.

%%%%%%%%%%%%%%%%%%%%%%%%%%%%%%%%%%%%%%%%%%%%%%
%%%%%%%%%%%%%%%%%%  %%%%%%%%%%%%%%%%%%%%%%%%%%
\section{Conservation Equation}\label{SecQE}
The off-shell invariance of the action $S=S_0+S_{\text{HS-$\S$~cubic 
interactions}}$ as constructed in \textsection \ref{CM}, up to order $g$ can be 
expressed in terms of a set of Bianchi identities. Using them together with the 
on-shell equations of motion of $\S$ we can show that the canonical higher spin 
supercurrent multiplet satisfies the following conservation equation:
\bea{l}
\Dd^{\ad_{k+1}}\J_{\a(k+1)\ad(k+1)}=\tfrac{1}{(k+1)!}\Dd^2\D_{(\a_{k+1}}\T_{\a(k))\ad(k)}\n\label{ce}~.
\eea
One can show that expressions (\ref{supercurrent}) and (\ref{supertrace}) 
automatically satisfy (\ref{ce}), given the free theory 
equation of motion $\D_{\a}\S=0$. Similarly, the minimal higher spin supercurrent multiplet (\ref{JminF}) satisfies
the conservation equation
\bea{l}
\Dd^{\ad_{k+1}}\J^{\textit{min}}_{\a(k+1)\ad(k+1)}=0\n\label{mince}~.
\eea
We can use this property to find a simpler expression for $\J^{\textit{min}}_{\a(k+1)\ad(k+1)}$.
Using as an ansatz the structure that appears in (\ref{JminF}) we can write
\bea{l}
\J^{\textit{min}}_{\a(s)\ad(s)}=\sum_{p=0}^{s}~a_{p}~\pa^{(p)}\S~\pa^{(s-p)}\Sd~+~\sum_{p=0}^{s-1}~b_{p}~\pa^{(p)}\Dd\S~\pa^{(s-p-1)}\D\Sd\n
\eea
and now we impose on this quantity two necessary conditions, reality and on-shell
conservation equation (\ref{mince}). The reality of $\J^{\textit{min}}_{\a(s)\ad(s)}$ gives the constraints
\bea{l}\n\label{reality}
a_{p}=a_{s-p}^{*}~,~p=0,\dots,s\sn~,~\\
b_{p}=b_{s-p-1}^{*}~,~p=0,\dots,s-1\sn~.
\eea
The conservation of $\J^{\textit{min}}_{\a(s)\ad(s)}$ (using $\D_{\a}\S=0$) gives:
\bea{l}
b_{p}~\frac{p+1}{s}-i~a_{p}~\frac{s-p}{s}=0~,~p=0,\dots,s-1~.\n\label{conservation}
\eea
These two constraints (\ref{reality}, \ref{conservation}) are enough to fix coefficients $a_{p},~b_{p}$ up to a real proportionality constant $c$
\bea{l}\n
a_{p}=c~i^{s}~(-1)^{p}~\binom{s}{p}^2~,~p=0,\dots,s~,~\sn\\
b_{p}=c~i^{s+1}~(-1)^{p}~\binom{s}{p}^2~\frac{s-p}{p+1}~,~p=0,\dots,s-1~.\sn
\eea
The overall constant of proportionality $c$, can be fixed by comparing with (\ref{JminF}). The result is
\bea{l}
\J^{\textit{min}}_{\a(s)\ad(s)}=-~\frac{(-i)^{s}}{\binom{2s+1}{s+1}}\sum_{p=0}^{s}~(-1)^{p}~\binom{s}{p}^2\left\{~\pa^{(p)}\S~\pa^{(s-p)}\Sd
~+~i~\left(\frac{s-p}{p+1}\right)~\pa^{(p)}\Dd\S~\pa^{(s-p-1)}\D\Sd\right\}\n\label{JminCon}~.~~
\eea

%%%%%%%%%%%%%%%%%%%%%%%%%%%%%%%%%%%%%%%%%%%%%%
%%%%%%%%%%%%%%%%%%  %%%%%%%%%%%%%%%%%%%%%%%%%%
\section{Higher spin supercurrents via Complex linear - Chiral Duality}\label{Duality}
We have already noticed that (\emph{i}) the complex linear supermultiplet 
couples only to half-integer superspin supermultiplets $(s+1,s+1/2)$, like the
chiral superfield does and (\emph{ii}) the corresponding minimal higher spin supercurrents for 
the complex linear superfield and the chiral superfield have many similarities. In order to find what is their 
precise connection, we perfom the well known duality procedure that maps one to 
the other.

As a starting point, we consider the auxiliary action
\bea{l}
S=-\int d^8z ~\bar{\s}~\s~+~\int d^8z ~\Phi~\s~+~\int d^8z ~\bar{\Phi}~\bar{\s}~
+~g\int d^8z \sum_{s=0}^{\infty}H^{\a(s)\ad(s)}\J^{\textit{min}}_{\a(s)\ad(s)}\n\label{AuxAction}
\eea
where $\s$ is an unconstrained, complex, scalar superfield, $\Phi$ is a chiral 
superfield and $\J^{\textit{min}}_{\a(s)\ad(s)}$ is given by (\ref{JminCon}) 
with $\S$ and $\Sd$ replaced by $\s$ and $\bar{\s}$ respectively. It is 
straightforward to see that once we integrate out $\Phi$,
$\s$ is promoted to a complex linear superfield $\S$ and we recover 
$S_{0}+S_{\text{HS-$\S$ minimal cubic interactions}}$ with the correct higher 
spin supercurrents. Now, if we instead integrate out $\s$, we get
\bea{l}
S=\int d^8z ~\bar{\Phi}~\Phi~+~g\int d^8z \sum_{s=0}^{\infty}H^{\a(s)\ad(s)}\J^{\textit{min}}_{\a(s)\ad(s)}\Bigr|_{\substack{\s=\bar{\Phi}}}\n~.
\eea
However, the quantity $\J^{\textit{min}}_{\a(s)\ad(s)}\Bigr|_{\substack{\s=\bar{\Phi}}}$
can be written as
\bea{ll}
\J^{\textit{min}}_{\a(s)\ad(s)}\Bigr|_{\substack{\s=\bar{\Phi}}}&=(-1)^{s+1}~\frac{(-i)^{s}}{\binom{2s+1}{s+1}}\sum_{p=0}^{s}(-1)^{p}
~\binom{s}{p}^2\left\{~\pa^{(p)}\Phi~\pa^{(s-p)}\bar{\Phi}
~+~i~\left(\frac{s-p}{p+1}\right)~\pa^{(p)}\D\Phi~\pa^{(s-p-1)}\Dd\bar{\Phi}\right\}\\
&=(-1)^{s+1}\J^{\textit{min}~(\Phi)}_{\a(s)\ad(s)}\n\label{clcJ}
\eea
where $\J^{\textit{min}~(\Phi)}_{\a(s)\ad(s)}$ is the minimal higher spin 
supercurrent constructed out of a chiral superfield \cite{hsch1,hsch2}.

The result of the duality transformation is a chiral theory with cubic interactions
to higher spin supermultiplets, given by
\bea{l}
S=\int d^8z ~\bar{\Phi}~\Phi~+~g\int d^8z 
\sum_{s=0}^{\infty}~(-1)^{s+1}~H^{\a(s)\ad(s)}\J^{\textit{min}~(\Phi)}_{\a(s)\ad(s)}\n~.
\eea
Notice that there is a difference in the coupling constant or \emph{charge} that 
controls the cubic interactions with higher spin supermultiplets.
We started by fixing all the higher spin charges of the complex linear theory to 
be the same [$g$] and we found that the corresponding charge of the dual chiral 
theory is spin dependent [$(-1)^{s+1}g$] and alternates in sign between even and 
odd values of $s$. For odd values of $s$ both theories have the same sign charge and 
for even values of $s$ they have opposite sign charge.

It is known that for the case of coupling to the vector multiplet $(s=0)$, the 
chiral and complex linear superfield have opposite sign charge. 
This can be easily understood be observing the
neutrality of the $\Phi~\s$ term in (\ref{AuxAction}). Our analysis indicates that this 
behavior extents to arbitrary high spin supermultiplets $(s+1,s+1/2)$ with even 
$s$. For the supergravity case $(s=1)$ and all higher spin 
supermultiplets $(s+1,s+1/2)$  with $s$ odd, both charges have the same sign. We remind the reader that the 
highest propagating spin of the $(s+1,s+1/2)$ supermultiplet is $j=s+1$, hence 
it would be interesting to determine whether this result has a connection with 
the fact that odd $j$ spins can have repulsive interactions and even $j$ spins 
have only attractive interactions \cite{AtrRep}.
 
%%%%%%%%%%%%%%%%%%%%%%%%%%%%%%%%%%%%%%%%%%%%%%
%%%%%%%%%%%%%%%%%%  %%%%%%%%%%%%%%%%%%%%%%%%%%
\section{Higher spin component currents for the complex linear superfield}\label{SecComp}
To conclude our analysis, we would like to extract the higher spin component 
currents contained inside the supercurrent multiplet.
The easiest way to identify them, is through their conservation equation. For 
this reason, we will project the superspace conservation equation (\ref{mince})
to components.

First of all notice that, due to (\ref{mince}) and the reality of 
$\J^{\textit{min}}_{\a(s)\ad(s)}$, the entire superfield 
satisfies a spacetime conservation equation
\bea{l}
\pa^{\a_{s}\ad(s)}\J^{\textit{min}}_{\a(s)\ad(s)}=0\n
\eea
hence all of its components will be conserved. However, because of (\ref{mince}) 
not all of these components are independent.
The independent ones are:
\bea{l}
\J^{\textit{min}~(0,0)}_{\a(s)\ad(s)}\sim\J^{\textit{min}}_{\a(s)\ad(s)}\Bigr|_{\substack{\th=0 
\vspace{0.4ex}\\ \thd=0}}\n\label{hsc1}
~,~\vspace{1ex}\\
\J^{\textit{min}~(1,0)}_{\a(s+1)\ad(s)}\sim\tfrac{1}{(s+1)!}\D_{(\a_{s+1}}\J^{\textit{min}}_{\a(s))\ad(s)}
\Bigr|_{\substack{\th=0 \vspace{0.4ex}\\ \thd=0}}\n\label{hsc2}~,~\vspace{1ex}\\
\J^{\textit{min}~(1,1)}_{\a(s+1)\ad(s+1)}\sim\tfrac{1}{(s+1)!(s+1)!}\left[\D_{(\a_{s+1}},\Dd_{(\ad_{s+1}}\right]
\J^{\textit{min}}_{\a(s))\ad(s))}\Bigr|_{\substack{\th=0 \vspace{0.4ex}\\ 
\thd=0}}\n\label{hsc3}~.
\eea
and they are all conserved on-shell.

The first one (\ref{hsc1}) is a bosonic, integer spin ($j=s$), R-symmetry 
current. Using (\ref{JminCon}) we find that it is proportional to
\bea{l}
\J^{\textit{min}~(0,0)}_{\a(s)\ad(s)}\sim-(-i)^{s}\sum_{p=0}^{s}~(-1)^{p}~\binom{s}{p}^2\left\{~\pa^{(p)}A~\pa^{(s-p)}\bar{A}~
+~i~\left(\frac{s-p}{p+1}\right)~\pa^{(p)}\bar{\psi}~\pa^{(s-p-1)}\psi\right\}\n\label{hsc1exp}
\eea
where $\S|=A$ and $\Dd_{\ad}\S|=\bar{\psi}_{\ad}$. The second one (\ref{hsc2}) 
is a fermionic, half-integer ($j=s+1/2$) current
\bea{l}
\J^{\textit{min}~(1,0)}_{\a(s+1)\ad(s)}\sim-(-i)^{s}\sum_{p=0}^{s}~(-1)^{p}~\binom{s}{p}^2~\frac{s+1}{s-p+1}
~\pa^{(p)}A~\pa^{(s-p)}\psi\n\label{hsc2exp}~.
\eea
The last one (\ref{hsc3}) is a bosonic, integer spin ($j=s+1$) current
\bea{l}
\J^{\textit{min}~(1,1)}_{\a(s+1)\ad(s+1)}\sim(-i)^{s}\sum_{p=0}^{s}~(-1)^{p}~\binom{s}{p}^2\left\{\vphantom{\frac12}
~i~\pa^{(p)}A~\pa^{(s+1-p)}\bar{A}
~-~i~[\tfrac{2s-p+1}{p+1}]~\pa^{(p+1)}A~\pa^{(s-p)}\bar{A}\right.\n\label{hsc3exp}\\
\hspace{39ex}+\left.\vphantom{\frac12}
[\tfrac{s+p+2}{p+1}]~\pa^{(p)}\bar{\psi}~\pa^{(s-p)}\psi~-~[\tfrac{s-p}{p+1}]~\pa^{(p+1)}\bar{\psi}~\pa^{(s-p-1)}\psi\right\}~.
\eea
All these currents are proportional to the ones constructed from the chiral 
theory in \cite{hsch1}.

%%%%%%%%%%%%%%%%%%%%%%%%%%%%%%%%%%%%%%%%%%%%%%
%%%%%%%%%%%%%%%%%%  %%%%%%%%%%%%%%%%%%%%%%%%%%
\section{Summary and discussion}\label{SecConclude}
Let us briefly summarize our results. This work is the continuation of 
\cite{hsch1} and it aims to investigate the possibility of
cubic interactions between a single, massless, complex linear supermultiplet and 
$4D,~\N=1$ higher spin supermultiplets. Using Noether's method we 
derive explicit expressions for the higher spin supercurrent multiplet that 
gives rise to such interactions. In \textsection\ref{dSigma} we give a first 
order transformation for the complex linear superfield. Its compatibility with 
the linearity condition will give a set of constraints to the transformation 
parameters, the solution of which gives structures similar to the zeroth order 
gauge transformations of some higher spin supermultiplets. The outcomes of 
Noether's procedure are:
\begin{enumerate}
\item[1.] Cubic interactions exist only for the half-integer superspin supermultiplets $(s+1,s+1/2)$.
\item[2.] The \emph{canonical} higher spin supercurrent multiplet $\{\J_{\a(k+1)\ad(k+1)},~\T_{\a(k)\ad(k)}\}$
includes the supercurrent (\ref{supercurrent}) and supertrace (\ref{supertrace}). 
Both of them generate the interaction terms (\ref{cubint}) and satisfy the superspace
conservation equation (\ref{ce}).
\item[3.] For every value of $k$, there exist an improvement term that will take
us from the \emph{canonical} supercurrent multiplet to the \emph{minimal} supercurrent
multiplet defined by $\T^{\textit{min}}_{\a(k)\ad(k)}=0$ which includes the minimal higher
spin supercurrent $\J^{\textit{min}}_{\a(k+1)\ad(k+1)}$ given by (\ref{JminF}, \ref{JminCon}).
It satisfies the superspace conservation equation (\ref{mince}) and the cubic interactions
it generates are (\ref{mincubint}).
\end{enumerate}
These results have similarities with the results presented in \cite{hsch1}, 
where the cubic interactions between higher spin and chiral supermultiplets are 
constructed. For that reason we check the well known duality between a 
chiral and complex linear theory in the presence of these higher spin 
cubic interactions. The result is:
\begin{enumerate}
\item[4.] The duality holds with an interesting twist. If the \emph{charge} that 
controls the cubic interaction of a complex linear superfield to the higher spin 
supermultiplet $(s+1,s+1/2)$ is $g$, then the corresponding \emph{charge} for 
the chiral theory is $(-1)^{s+1}g$. Therefore, for odd values of $s$ (such as 
supergravity, s=1) both the chiral and complex linear superfields have the same sign higher spin
charge. However, for even values of 
$s$ (such as the vector multiplet, s=0) they have opposite sign higher spin charge.
\end{enumerate}
Finally, we focus at the component level of the supercurrent multiplet and 
identify the various spacetime conserved higher spin currents. There are three 
higher spin currents:
\begin{enumerate}
\item[5.] There is a bosonic, spin $j=s$ current $\J^{\textit{min}~(0,0)}_{\a(s)\ad(s)}$ 
(\ref{hsc1},\ref{hsc1exp}) which corresponds to the $\th$ independent component of 
$\J^{\textit{min}}_{\a(k+1)\ad(k+1)}$. This current corresponds to an 
R-symmetry.
\item[6.] There is a second bosonic, spin $j=s+1$ current 
$\J^{\textit{min}~(1,1)}_{\a(s+1)\ad(s+1)}$ (\ref{hsc3},\ref{hsc3exp}) corresponding to the 
$\th\thd$ component.
\item[7.] There is a fermionic, spin $j=s+1/2$ current 
$\J^{\textit{min}~(1,0)}_{\a(s+1)\ad(s)}$ (\ref{hsc2},\ref{hsc2exp}), corresponding to the 
$\th$ component.
\end{enumerate} 
Notice that the bosonic currents have two independent contributions, one coming from the bosonic sector (complex scalar)
and another coming from the fermionic sector (spinor). These two contributions have been discovered independently
by studying non-supersymmetric theories \cite{current1, current2, current3}. However, the fermionic current has a single contribution
that depends on both the boson and the fermion, hence the discovery of such a higher spin current appears naturally in supersymmetric
theories. The first time this type of current appeared was in \cite{hsch1}, where the complex scalar
and spinor are defined as components of a chiral superfield. In this work we find the analogue expressions for
the case of using a complex linear superfield to define them. The results we get are proportional to the ones found in \cite{hsch1}.
and consistent with the duality transformation discussed in \textsection\ref{Duality}.

%%%%%%%%%%%%%%%%%%%%%%%%%%%%%%%%%%%%%%%%%%%%%%
%%%%%%%%% Acknowledgement %%%%%%%%%%%%%%%%%%%%
{\bf Acknowledgements}\\[.1in] \indent
K.\ K.\ wants to thank S.\ J.\ Gates Jr. and I.\ L.\ Buchbinder for usefull discussions.
P.\ K.\ wants to thank Michal Pri\v{s}egen and Samuel Moln\'{a}r for discussions. The work of K.\ K.\ and R.\ v.\ U.\ was supported by the grant P201/12/G028 of the Grant
agency of Czech Republic.

%%%%%%%%%%%%%%%%%%%%%%%%%%%%%%%%%%%%
%%%%%%%%% Bibliography %%%%%%%%%%%%%%%%%%%%

\end{document}